\documentclass[useAMS,usenatbib]{mn2e}
\usepackage{amsmath}
\usepackage{graphicx}

\usepackage{ enumitem, tabto, xfrac}
\usepackage[]{units}
\usepackage{tabu}
\usepackage{morefloats}
\usepackage{float}
\usepackage{amssymb}
\usepackage{xcolor}
\usepackage[normalem]{ulem}

\title[Limits on Foreground Subtraction from Chromatic Beam Effects in Global Redshifted 21~cm Measurements]{Limits on Foreground Subtraction from Chromatic Beam Effects in Global Redshifted 21~cm Measurements}
\author[T. J. Mozdzen, J. D. Bowman, R. A. Monsalve, and A. E. E. Rogers]{T. J. Mozdzen$^{1}$\thanks{E-mail:
tmozdzen@asu.edu (TJM)}, J. D. Bowman$^{1}$, R. A. Monsalve$^{1}$, and A. E. E. Rogers$^{2}$\\
$^{1}$School of Earth and Space Exploration, Arizona State University (ASU), Tempe, Arizona, 85287, USA\\
$^{2}$MIT Haystack Observatory, Massachusets Institute of Technology (MIT), Westford, Massachusetts, 01886, USA}
\begin{document}

\date{Accepted 2015 MMa DDa. Received 2015 MMr DDr; in original form 2015 MMo DDo}

\pagerange{\pageref{firstpage}--\pageref{lastpage}} \pubyear{2015}

\maketitle

\label{firstpage}

\begin{abstract}
\textcolor{black}{Foreground subtraction in global redshifted 21 cm measurements is limited by frequency-dependent (chromatic) structure in antenna beam patterns. Chromatic beams couple angular structures in Galactic foreground emission to spectral structures that \textcolor{black}{may not be} removed by smooth functional forms. We report results for simulations based on two dipole antennas used by the Experiment to Detect the Global EoR Signature (EDGES).  The  residual levels in simulated foreground-subtracted spectra are found to differ substantially between the two antennas, suggesting that antenna design must be carefully considered.  Residuals are also highly dependent on the right ascension and declination of the antenna pointing, \textcolor{black}{with RMS values} differing by as much as a factor of 20 across pointings.  For EDGES and other ground-based experiments with zenith pointing antennas, right ascension and declination correspond directly to the local sidereal time and the latitude of the deployment site, hence chromatic beam effects should be taken into account when selecting sites.    We introduce the ``blade'' dipole antenna \textcolor{black}{ and show, via simulations, that it has better chromatic performance than the ``fourpoint'' antenna previously used for EDGES.}  The blade antenna yields 1-5~mK residuals across the entire sky after a 5-term polynomial is removed from simulated spectra, whereas the fourpoint antenna typically requires a 6-term polynomial for comparable residuals.  For both antennas, the signal-to-noise ratio of recovered 21 cm input signals peaks for a 5-term polynomial foreground fit given realistic thermal noise levels.}  
\end{abstract}

\section{INTRODUCTION}

The epoch of reionisation (EoR) is the focus of significant theoretical and observational research efforts due to its importance in the understanding of cosmic evolution.  The time of its onset and duration are related to fundamental information about the first stars, galaxies, and accreting black holes, including mass, radiative output, and composition \citep{b6}. Theoretical models place the EoR between redshifts of \textcolor{black}{$20\gtrsim z \gtrsim 6$}, after which the intergalactic medium (IGM) is observed to be fully ionised \citep{b1, b3, b2}.  The duration of the EoR is difficult to predict theoretically \textcolor{black}{ \citep{b18,  b8a}} and observational constraints are just beginning to emerge \textcolor{black}{\citep{ b4c, b52, b70, b65, b01, b15a}}. 

At radio wavelengths, the 21~cm hyperfine transition of neutral hydrogen (rest frequency of 1420~MHz) provides a versatile signal for studying \textcolor{black}{the epochs of Cosmic Dawn and Reionization} by probing the temperature and ionisation state fraction of neutral hydrogen gas in the IGM. The detectable brightness temperature due to redshifted 21~cm emission or absorption from the early IGM is given by \citep{b5, b3b, b6, b7, b8} 
\begin{equation} 
\label{eq:sky_temp}
T_{\text{b}}(z) \approx 27x_{\text{HI}}\left(\frac{1+z}{10}\right)^\frac{1}{2}\left(1-\frac{T_{\text{CMB}}}{T_\text{S}}\right) \text{mK,}
\end{equation}
where $x_{\text{HI}}$ is the neutral fraction of the gas, $T_\text{S}$ is the ‘spin’ temperature that describes the relative populations of the ground and the hyperfine excited states, and $T_\text{CMB}$ is the temperature of the cosmic microwave background (CMB) radiation, all of which depend upon $z$ implicitly.

\begin{figure}
\centering
  \includegraphics{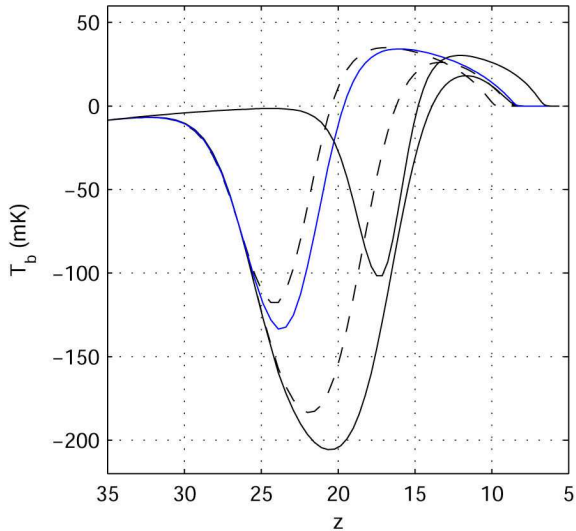}
  \caption{A sample of theoretical models for the 21~cm brightness temperature with various values for model parameters (courtesy \citealt{b8a}). The models predict the magnitude of the 21~cm signal to peak between 15 and 40~mK in the interval between redshifts 6 and 20.}
  \label{fig:Theory}
\end{figure}

The spin temperature amplitude is affected by UV radiation via the Wouthuysen-Field mechanism \citep{b9,b10}, collisions with IGM gas, and interactions with CMB photons. Thus $T_\text{b}$ can be either positive or negative depending upon the spin temperature relative to the CMB temperature.   The shape of the \textcolor{black}{$T_\text{b}$} vs redshift curve indicates the relative strength and timing of the early processes  \textcolor{black}{mentioned above} and  theoretical studies have varied the model parameters to produce families of resultant curves (see e.g. \citealt{b6a, b6, b12, b18, b19, b8a}).

There are two main approaches to studying the EoR at radio wavelengths. The first method attempts to detect the EoR statistically, primarily through the redshifted 21~cm fluctuation power spectrum, and eventually to image large structures directly. Efforts such as the Low-Frequency Array for Radio astronomy (LOFAR\footnotemark\footnotetext{www.lofar.org}, \textcolor{black}{\citealt{b14, b81}}), the Precision Array to Probe the Epoch of Re-ionization (PAPER\footnotemark\footnotetext{http://eor.berkeley.edu/}, \citealt{b01, b15a}),  the Murchison Widefield Array (MWA\footnotemark\footnotetext{www.mwatelescope.org}, \citealt{b16}; \citealt{b40}), the Square Kilometer Array (SKA\footnotemark\footnotetext{www.skatelescope.org}, \citealt{b17}), and the  Hydrogen Epoch of Reionization Arrays (HERA\footnotemark\footnotetext{http://reionization.org/}) are radio interferometers currently operating or in development that aim to recover the redshifted 21~cm power spectrum.

The second method aims to detect the global 21~cm signal through full sky observations using a single antenna. A global signal antenna will respond to a range of frequencies and $T_\text{b}(z)$ will correspond to the frequency range of redshifted 21 cm emissions. The predicted spectral signature is broadband between 50 and 200~MHz (30 $\leq$ $z$ $\leq$ 6), with a peak absolute amplitude between 10 and 200 mK dependent upon particular star formation model parameters chosen (see Fig. \ref{fig:Theory}, data from \citealt{b8a}).  Galactic and extragalactic continuum foregrounds from synchrotron and free-free emission are approximately four orders of magnitude larger, with typical sky temperatures of 250~K at 150~MHz away from the \textcolor{black}{Galactic Centre}.  The foregrounds generally exhibit smooth, power-law-like spectra that must be subtracted from observations to reveal the 21~cm signal.  

Several global 21~cm experiments are in progress using various radio receiver architectures and antenna design styles. The major efforts include the Experiment to Detect the Global EoR Signature (EDGES) \citep{b4c}, Dark Ages Radio Explorer (DARE) \citep{b23}, Broadband Instrument for Global Hydrogen Reionisation Signal (BIGHORNS) \citep{b46}, Shaped Antenna measurement of background Radio Spectrum (SARAS) \citep{b49, b49a}, Large-aperture Experiment to detect the Dark Age (LEDA) \citep{b21}, and Sonda Cosmol\'{o}gica de las Islas para la Detecci\'{o}n de Hidr\'{o}geno Neutro (SCI-HI) \citep{b62}.  Several studies have addressed experimental calibration issues \textcolor{black}{\citep{b4d, b49, b21}} as well as the frequency dependence of the antenna beams \textcolor{black}{\citep{b60,b47,b21}}.  

In this study we focus on antenna beam effects in the detection of the global 21~cm signature in the range of 13.2 $> z >$ 6.4 (100 $< \nu <$ 190~MHz) \textcolor{black}{across sky position in right ascension and declination, which we map to the local sidereal time (LST) and latitude of the experiment deployment site}.  The antenna is a critical element \textcolor{black}{of the EDGES system and since it is not embedded in an array, its beam} cannot be readily calibrated as part of the observing program  (see \citealt{b21}). The antenna must be designed carefully and modeled accurately {\it a priori} to compensate for its characteristics during observations.  

The primary design objective for the antenna \textcolor{black}{beam is a directivity pattern that varies smoothly in frequency}.  Chromatic antenna beams are undesirable because they can couple the relatively large angular fluctuations in the Galactic foreground into \textcolor{black}{spectral structures that may confuse global 21~cm signatures}.  

\textcolor{black}{\cite{b47} have evaluated the chromatic effects of the ionosphere with the LOFAR low frequency antenna beam, and \cite{b21} investigated the chromatic effects of a detailed sky foreground model with analytical forms of the LEDA \textcolor{black}{dipole} beam. Here, we compare the chromatic effects of two dipole-based antennas and an idealized reference antenna, over deployment latitude and LST, in the context of the EDGES project. We isolate the effects of the antenna beams by ignoring ionospheric effects and} \textcolor{black}{by adopting a power-law model for the foreground emission.} In Section~\ref{sec:methods} we describe the antennas and the method used to calculate their response. In Section~\ref{sec:results} we discuss the results of our simulations and conclude in Section~\ref{sec:conclusion} with a summary of key findings and a discussion of implications and potential future investigation paths.

\section{METHODS}
\label{sec:methods}

We base the instrument model in our simulations on the EDGES project (\citealt{b4a, b4b, b4c, b4d}). It employs a broadband dipole-like antenna sensitive to wavelengths between 3 and 1.6~meters (100 - 190~MHz). The antenna is connected to a radio receiver that amplifies and conditions the signal before passing it to a digital spectrometer that samples the spectrum with 6~kHz resolution. The receiver utilizes laboratory calibration prior to deployment, augmented with a three-position hot/cold calibration switching scheme during operations \citep{b4d}, to achieve an accuracy of $\sim$0.01 \% in measured antenna temperature as a function of wavelength. The impedance match of the antenna connection to the receiver is measured \textit{in situ} by periodically switching a Vector Network Analyzer (VNA) into the electrical path. Although the EDGES calibration scheme is sufficient to correct for undesirable electronic effects in the measured spectrum, it does not compensate for chromatic beam effects. 
\subsection{Antennas}
\begin{figure}
\centering
  \includegraphics{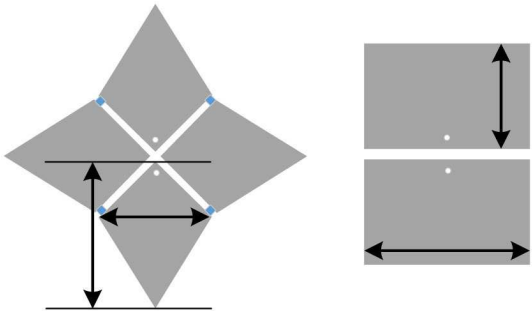}
  \caption{Fourpoint antenna (left) and blade antenna (right) are shown in a top view with dimension-indicating arrows to denote the individual panel widths and lengths listed in Table \ref{tab:antennas}.}
  \label{fig:Fourpoint_and_Blade}
\end{figure}

\begin{figure*}
  \begin{minipage}[b]{0.99\linewidth}
  \centering
  \includegraphics{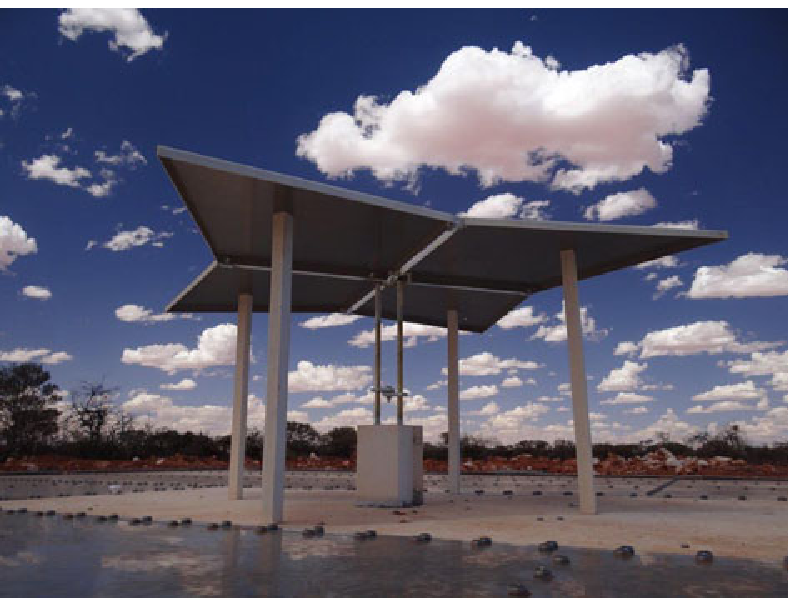}
  \includegraphics{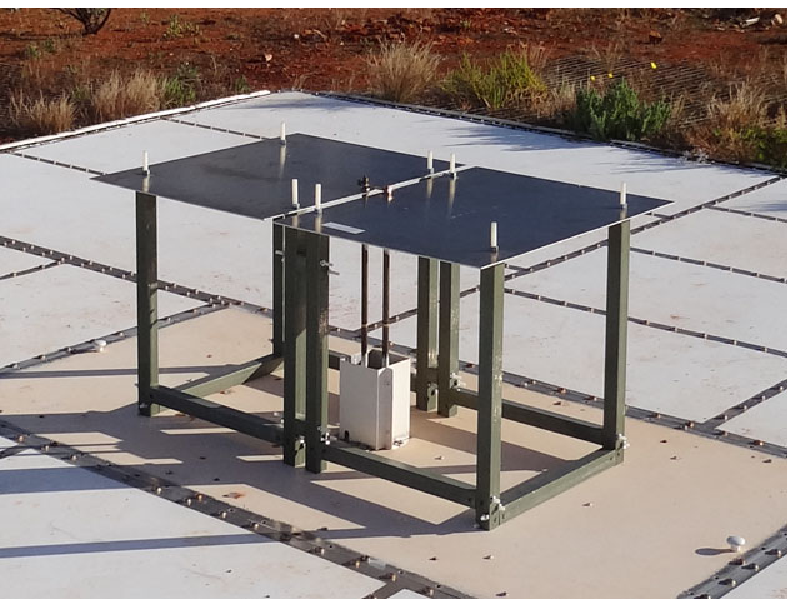}
  \end{minipage}
 \caption{(left) Photograph of the fourpoint antenna as deployed by EDGES in 2015. The fourpoint design has a downward pointing rim (1.8~cm) on the perimeter of each panel and uses discrete capacitors between the panels at the outer edge as well as a tuning capacitor half-way up the tubes which are part of the Roberts balun. (right) \textcolor{black}{Photograph} of the blade antenna which does not use inter-panel capacitors, a balun tuning capacitor, nor a perimeter rim. (common) Both designs use fiberglass support tubes, four for the fourpoint and eight for the blade.  Both antennas use a tuning capacitor on the top of the panels between the balun tubes to improve impedance matching. Surrounding the tubes at the base, a short rectangular enclosure shields against vertical currents in the tubes.}
  \label{fig:antenna_images}
\end{figure*}

In this study, we analyse three horizontal planar dipole-like antennas placed over a ground plane.  Each antenna is tuned to respond in the EDGES band.  The three antenna types are: 1) The EDGES ``fourpoint'' antenna deployed to date, which is based on the fourpoint design of \citet{b24}; 2) a ``blade'' design that shows superior beam qualities in simulations and is being considered as a potential successor to the fourpoint design; and 3) an analytic \nicefrac[]{1}{2}-$\lambda$ wire dipole, which is included as an analytic comparison.   The fourpoint and blade antennas are shown in Figs.~\ref{fig:Fourpoint_and_Blade}~-~\ref{fig:antenna_images} and Table \ref{tab:antennas} summarizes the design and model parameters of the antennas.

Numerical time-domain electromagnetic simulations were performed using CST (Computer Simulation Technology) Microwave Studio for the fourpoint and blade antennas.  All of the antenna components were simulated except for fiberglass support legs and cable connectors. Metal structures were modeled with their actual dimensions and thicknesses, but ground planes were modeled as infinite metal sheets. \textcolor{black}{We found that the choice of CST settings can affect the chromaticity of the modeled beams by introducing numerical artifacts from rounding and precision errors.   In particular, we needed to carefully select transient power dissipation thresholds and mesh grid resolutions in order to minimize such artifacts while maintaining efficient processing times.  To fine-tune our CST settings, we performed a convergence study in which we perturbed the physical antenna dimensions (by of order 1\%) in our CST models while adjusting simulator settings until the resulting outputs converged to small variations.  We found this test for convergence of results, using nearly identical antenna models, to be a good probe of the level of numerical artifacts in the antenna simulations.}  

\textcolor{black}{Our final antenna simulations were performed using CST settings that led to no more than 0.02 dB variations in reflection coefficients between perturbed antenna models over the frequency range of interest and yielded no more than 1.5 mK variations in foreground-subtracted residuals using a 5-term polynomial fit after our full analysis for antennas modeled at -26 degrees latitude with LST of 4 h.} The mesh cell counts were 13~million cells for the fourpoint antenna and 6~million cells for the blade antenna. Simulations required approximately 20~min. for the blade and 40~min. for the fourpoint antenna when using an NVIDIA \textcolor{black}{M2090} GPU accelerator. Peak memory requirements were modest at less than 8~GB.   We briefly describe each antenna below.

\begin{table*}
   \caption{Antenna Features (refer to Figs. \ref{fig:Fourpoint_and_Blade}-\ref{fig:antenna_images})}
    \label{tab:antennas}
     \begin{tabular} {@{}lc@{\hspace{2.5em}}c@{\hspace{0.7em}}c@{\hspace{0.7em}}c@{\hspace{0.7em}}c@{}}
   \hline
   & \multicolumn{2}{c}{\underline{3 dB Beamwidth at 150 MHz}}&Height above& Panel Width& Panel Length\\
   Antenna&  $\phi$=0$^{\circ}$ &$\phi$=90$^{\circ}$ &ground plane& $\bot$  to excitation axis& $\parallel$ to excitation axis\\
    &    (Degrees)&  (Degrees)&(cm)&  (cm)&  (cm)\\
   \hline
   \nicefrac[]{1}{2}-$\lambda$  wire dipole & 70 &   111 & 42.5&N/A&84.9$^a$\\
   Fourpoint                           & 66 &   105 & 51.8&53.0&69.7\text{ } \\
   Blade                                 & 72 &   110 & 52.0&62.6&48.8\text{ } \\
   \hline
\textcolor{black}{$^a$dipole full length}
\end{tabular}
\end{table*}
{\bf Fourpoint antenna:}  The fourpoint antenna uses four diamond-shaped panels \textcolor{black}{arranged in a } planar structure.  One pair of opposing panels is electrically active, while the other pair serves as a parasitic capacitance via a vertical rim along the panel's perimeter, to enhance both the beam's symmetry and the antenna's impedance match to the receiver (Fig. \ref{fig:antenna_images}). A Roberts transmission-line balun \citep{b30} is used to transition from the panels to the receiver.  Discrete tuning capacitors located at roughly the middle of the Roberts balun and near the edges of the panels, along with a capacitive top plate above the central region of the antenna, improve the impedance match of the antenna to the receiver.  

EDGES has deployed this style of antenna at the Murchison Radio-astronomy Observatory (MRO) in Western Australia for several observing seasons. A version of this style without the Roberts balun provided the data used to set a lower limit on the duration of the reionisation epoch \citep{b4c}. 

\begin{figure}
\centering
  \includegraphics{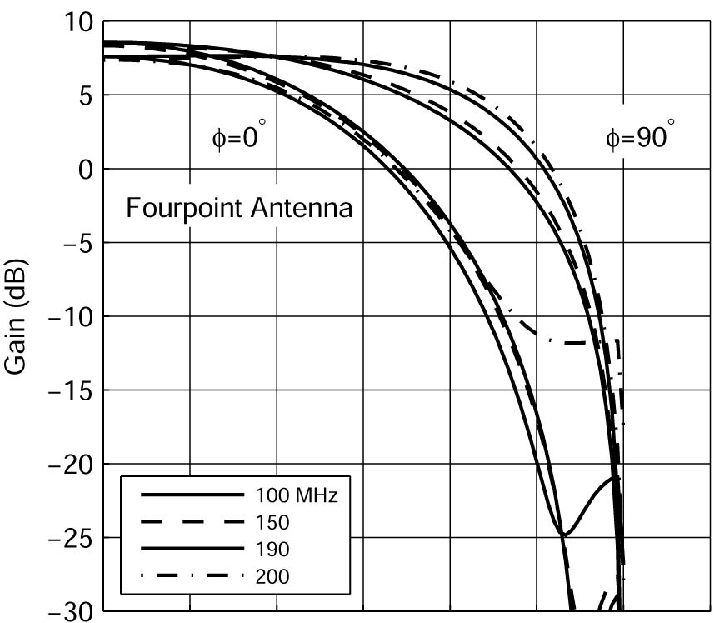}
  \includegraphics{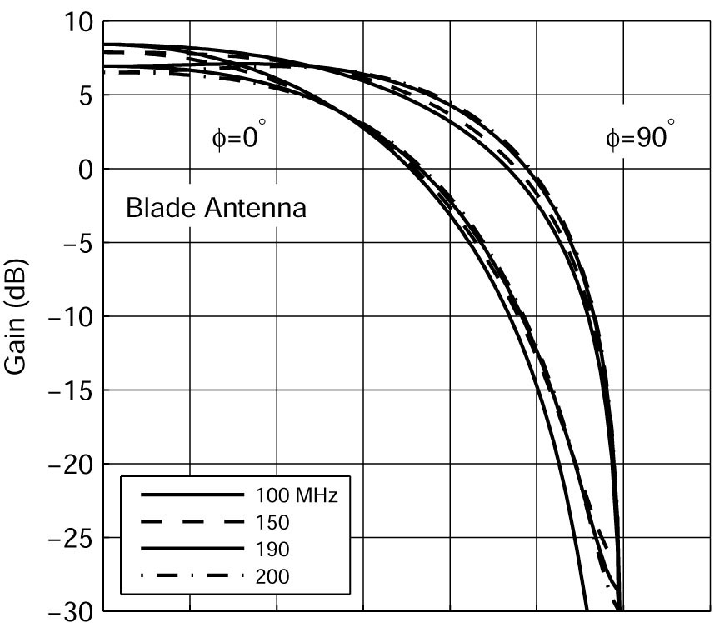}
  \includegraphics{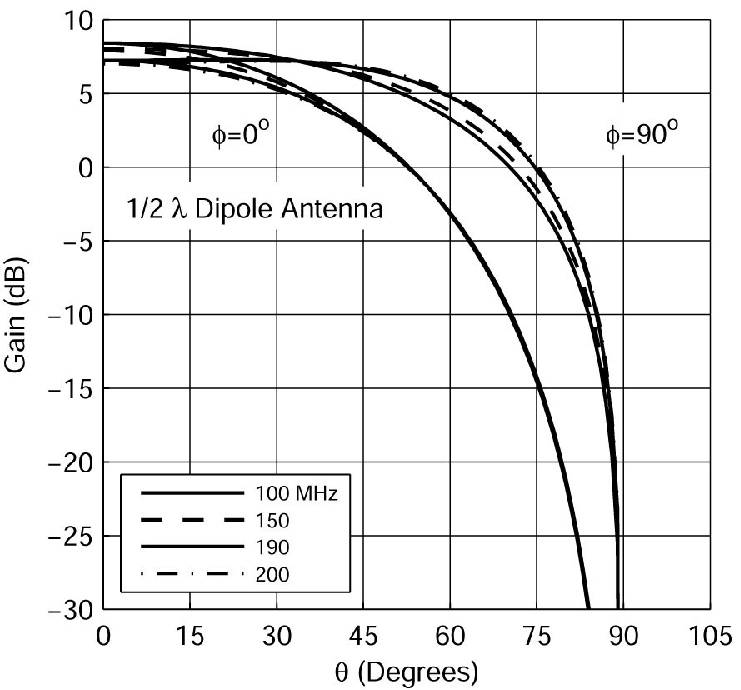}
  \caption{Cross-sections of simulated beam patterns for (top) the fourpoint antenna, (middle) the blade antenna, and (bottom) the analytic   \nicefrac[]{1}{2}-$\lambda$  wire antenna  at $\phi= 0^\circ$ (E-plane) and 90$^\circ$ (H-plane) for various frequencies. The active excitation axis of each antenna is defined as  $\phi=0^\circ$.  The horizon response ($\theta \approx 90^\circ$) is generally largest and most frequency-dependent for the fourpoint. Non-smooth beam changes are very small and not visible at these scales. Derivative plots are best used to detect these changes and are discussed in section \ref{subsec:origin}.}
  \label{fig:cross-sections}
\end{figure}

{\bf Blade antenna:}  The blade-shaped antenna is simpler than the fourpoint design, because it uses two flat rectangular panels (no rim on the perimeter), a top capacitor, the Roberts Balun, and a small shield at the bottom (Fig. \ref{fig:antenna_images}). There are no inter-panel tuning capacitors nor a balun tuning capacitor. The beam has less variation with frequency than the fourpoint design as can be seen in Fig. \ref{fig:cross-sections}. The  \textcolor{black}{ground plane consists of} a 5~m $\times$ 5~m solid aluminum plate underneath the antenna, with four wire mesh extensions (each 2~m $\times$ 5~m) which form a `plus-shaped' ground plane.  This antenna \textcolor{black}{was deployed in the field for the first time in July 2015}. 

{\bf Ideal \nicefrac[]{1}{2}-$\lambda$  wire antenna:} 
\label{subsec:half_wavelength}
The theoretical reference beam in our study is that of an  \textcolor{black}{infinitesimally} thin \nicefrac[]{1}{2}-$\lambda$  horizontal wire dipole antenna placed \textcolor{black}{a quarter wavelength, at a reference wavelength $\lambda_0$,} above an infinite ground plane.  \textcolor{black}{It has a near ideal beam shape that} will be explained in subsequent sections. The equation for the beam can be derived from the finite vertical dipole wire antenna \citep{b29} rotated on its side with a ground plane serving as the array factor and is given by:
\begin{equation} 
\label{eq:DipoleLong_beam}
B_{\nicefrac[]{1}{2}\text{-}\lambda} =  \left[\frac{\cos(\frac{\pi L}{\lambda} \cos\theta') - \cos(\frac{\pi L}{\lambda })} {\sin\theta'}\right]^2 \sin^2\left(2\pi h\cos\theta\right),
\end{equation} 
where $\theta'=\cos^{-1}(\sin\theta \sin\phi)$, $\theta$ and $\phi$ are the spherical angle coordinates with $\theta=0$ aligned to principal axis orthogonal to the ground plane and $\phi=0$ aligned to the active excitation axis of the antenna, $L$ is the full length of the antenna,  and $h$ is  the \textcolor{black}{height of the antenna above the ground plane in reference wavelengths.}

Figure \ref{fig:cross-sections} shows the beam pattern variation of each antenna model with frequency for $\phi$ = 0$^\circ$ and 90$^\circ$ and  illustrates the frequency-dependent variations in these antennas over the wide range of operating frequencies as well as the angular variation with elevation.  As common to most dipole-based antennas, all three antennas have large primary beams (fields of view) that span up to $\sim$$110^{\circ}$ FHWM.  The beam patterns of the antennas vary smoothly with respect to angle and with respect to frequency as viewed in Fig. \ref{fig:cross-sections}, but for global 21~cm measurements, smaller features that vary by less than 1\% \textcolor{black}{that are not readily apparent in the figure,} can cause undesirable chromatic effects.  

When the height of an antenna above the ground plane becomes larger than a quarter wavelength, structures in the beam near the zenith and sidelobes begin to form at that frequency. To avoid this unwanted structure, one would ideally place the antenna at a quarter wavelength above the ground plane for the highest frequency (shortest wavelength), ensuring that the height remains under a quarter wavelength for all frequencies.  However, the height of the fourpoint antenna is chosen as a compromise to optimize the impedance match while keeping the beam shape smooth.  For the fourpoint and blade antennas, we use a height above the ground plane that is equivalent to a quarter wavelength at 150~MHz.  This produces a minor ($<$ 0.2~dB) local minimum at the zenith for frequencies above 150~MHz.  The theoretical \nicefrac[]{1}{2}-$\lambda$  wire antenna was placed slightly lower above the ground plane, comparable to a quarter wavelength at 176.7~MHz, in order to create local minima of similar amplitude to those from the fourpoint antenna.  

\subsection{Sky Model}

\begin{figure}
\centering
  \includegraphics{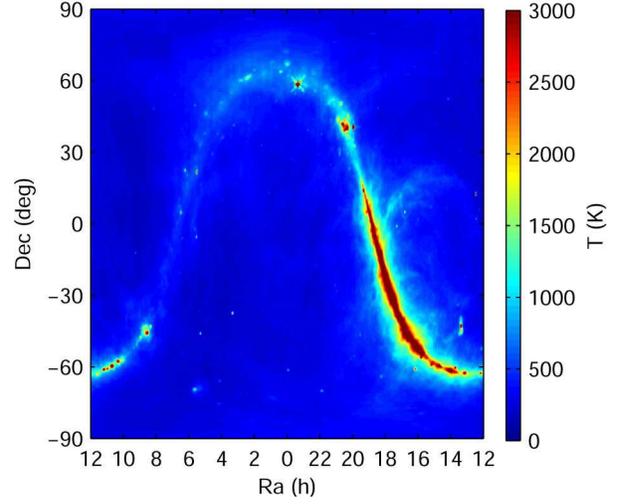}
  \caption{Sky model at 150~MHz extrapolated from the 408~MHz sky map of \citealt{b31}.}
  \label{fig:Haslam_sky_plot}
\end{figure}

The sky foreground model in our simulations is based on the \citet{b31} map at 408~MHz as shown in Figure \ref{fig:Haslam_sky_plot} with approximately 0.6$^\circ$ angular resolution. We model $T_\text{sky}$ as a simple power law by scaling the sky temperature as a function of frequency according to: 
\begin{equation} 
\label{eq:Tsky}
T_{\text{sky}}(\nu,\boldsymbol{\zeta}) =T_\text{Haslam}(\boldsymbol{\zeta}) \left(\frac{\nu}{408\, \mathrm{ MHz}}\right)^{-\beta}, 
\end{equation}
where $\nu$ is frequency, $\boldsymbol{\zeta}$ is the sky coordinate vector, and $\beta$ is the power law spectral index which we set to 2.5 \citep{b4a}.  Our sky model does not contain the EoR signal, nor does it include ionospheric distortions (see \citealt{b47, b48, b58} for a discussion of ionospheric effects).   \textcolor{black}{Earlier studies have used \textcolor{black}{3-parameter} sky models \citep{b45}, and more recently have considered complex sky models of up to 6$^\text{th}$ order (7 terms) \citep{b21}}. We have not included this structure in our sky model in order to maintain simplicity.  Using a simple spectral model is sufficient here because we find (discussed in detail below) that the chromatic effects of our modeled antenna beams produce \textcolor{black}{a larger magnitude of spectral structure at high-orders} and, hence, will require similarly high-order model fits.  \textcolor{black}{We assume that} any complicated spectral structure inherent to the sky will be removed along with the chromatic beam effects.

\subsection{Measurement Equation}

Assuming an ideal receiver, we calculate the simulated antenna temperature using:

\begin{equation} 
\label{eq:T_antenna}
T_\text{ant}(\nu,\hat{\boldsymbol{n}}) = \int_{\Omega}T_\text{sky}(\nu,\boldsymbol{\zeta}) B(\nu,\boldsymbol{\zeta},\hat{\boldsymbol{n}}) \mathrm{d}\Omega ,
\end{equation}
where $\hat{\boldsymbol{n}}(\alpha, \delta, \psi)$ is the antenna pointing vector and is a function of the right ascension and declination ($\alpha, \delta$) of the primary beam axis that is perpendicular to the antenna ground plane. The orientation of the antenna along its third degree of freedom is specified by the angle of the active antenna axis east of north and is labeled $\psi$. $B(\nu, \boldsymbol{\zeta},\hat{\boldsymbol{n}})$ is the chromatic beam for a given antenna, pointing, and orientation, and is normalized to a unit integral at each frequency.   

For an antenna on the surface of the Earth pointed toward zenith, the pointing declination of its primary axis corresponds to the latitude of the antenna's deployment site, and the pointing right ascension corresponds to the LST at the site. As the earth rotates, the antenna pointing changes direction, altering the mapping of its beam pattern onto sky coordinates. 

\subsection{Figure of Merit}
We define the figure of merit (FoM) for assessing the significance of chromatic beam effects as the RMS residuals to a least squares best fit model:

\begin{equation}
\label{eq:FoM}
\mathrm{FoM(\hat{\boldsymbol{n}})} = \sqrt{\left < \left [T_\text{ant}(\nu,\hat{\boldsymbol{n}}) - T_\text{model}(\nu,\hat{\boldsymbol{n}}) \right]^2 \right >_{\nu}}.
\end{equation}

Chromatic antenna beams that couple little structure into the measured spectrum will produce small FoM values.   A good antenna for the global redshifted 21~cm measurement would yield residuals well below the expected 21~cm signal strength of 10 - 40~mK in the frequency range of 100 - 190~MHz while minimizing the number of free parameters in the model.  For our model equation, we use an N-term power-law polynomial given by:

\begin{equation} 
\label{eq:Power_Law_2}
T_\text{model} = \sum_{i=0}^{N-1} a_i\left(\frac{\nu}{\nu_0}\right)^{-2.5+i}.
\end{equation} 
This polynomial form generally produces good fits at low-orders in existing EDGES measurements.  The antenna temperature produced by our sky model and an ideal beam that is invariant with frequency could be fit exactly with only one term:  $T_\text{model} = a_{0} \nu^{-2.5}$, but as we will show, realistic antennas require that N$\approx$\textcolor{black}{5}.  

In order to simultaneously remove the foreground and estimate the global 21~cm signal, the number of terms in $T_\text{model}$ would have to be augmented with at least one more term to solve for the global 21~cm signature.  \textcolor{black}{With a 21~cm model term added to the fitting equation, we define the signal to noise ratio (SNR) to be}
\begin{equation} 
	\label{eq:SNR}
	\textcolor{black}{SNR = \frac{T_{peak}}{[(\sigma_{rms}^2+\sigma_T^2)(\bold{X}^t\bold{X})^{-1}_{21~cm}]^{1/2}},}
\end{equation}
\textcolor{black}{where $T_{peak}$ = 20 mK is a nominal value of the global 21~cm peak signal strength, $\sigma_{_T}$ is  the typical thermal noise expected from averaging multiple observations, $\sigma_{rms}=\mathrm{FoM}$ is the rms error of the residuals to the fit, and $\sigma_{rms}^2(\bold{X}^t\bold{X})^{-1}_{21~cm}$ is the 21~cm \textcolor{black}{auto-covariance} term in the covariance matrix. The design matrix $\bold{X}$ has N columns (equal to the number of terms in the fitting equation) and one row for every discrete frequency (91 in our case).}

\textcolor{black}{We choose a Gaussian shaped 21~cm fitting term inspired by the emission and absorption signatures in Fig. \ref{fig:Theory} in the form of}
\begin{equation} 
\label{eq:EOR}
T_{21}=a_{_{N+1}}e^{-\frac{(\nu-\nu_0)^2}{2\sigma^2}},
\end{equation} 
\textcolor{black}{where $\nu_0$ = 150~MHz and $\sigma$  is related to the Full Width Half Maximum (FWHM)  by $\sigma = \text{FWHM}/(2\sqrt{2\text{ln}2})$. The noise estimate is based upon averaging a week of observation data using spectral channels of 1.0~MHz and assumes the nosie is Gaussian and spectrally flat. Typical values are under 3~mK.}

\textcolor{black}{Properly accounting for the effects of covariance between the chromatic foregrounds and the signal parameters requires one to marginalize over foreground parameters. We use the SNR as defined in Eq. \ref{eq:SNR} as an approximate method to account for degeneracies arising from covariance between foreground and signal parameters.} \textcolor{black}{Using a higher-order foreground model lowers the RMS error, but raises the covariance of the 21~cm signal estimate which lowers the SNR value, indicating there should also be an upper limit to the number of terms one should use to remove the foreground and this will be explored in Section \ref{sec:results}. Since  $(\bold{X}^t\bold{X})^{-1}_{21~cm}$ depends only upon the design matrix X, which is not a function of the antenna, the antenna location, or the antenna pointing, we will focus on our FoM in Section 3 as a sufficient metric for analyzing differences between antennas and pointings.} 

\section{RESULTS}
\label{sec:results}
In this section we present the FoM for each of the three antenna beam models.   We investigate antenna pointings spanning the entire sky in right ascension and declination.  In order to facilitate interpretation for ground-based drift-scan experiments, where the pointing parameters are typically dictated by site location and observing time, we label the plots in this section with latitude (declination) and LST (right ascension), although we note that the results are equally valid for proposed space-based experiments such as the Dark Ages Radio Explorer (DARE,  \citealt{b23}) that would more naturally label the pointings with right ascension and declination.

Figure~\ref{fig:pcolor_poly6_M} shows the FoM as a function of latitude and LST for all three antennas using using N=6 polynomial terms. The resolution of the map is 1$^\circ$ in latitude and 4 minutes in LST, yielding a $(181\times360)$ data array. As evident in the figure, the fourpoint antenna yields the worst performance with typical FoM values ranging between 20 and 100~mK.  Nevertheless, there are regions of the sky where the FoM reaches 4~mK and observations with a fourpoint antenna can still enable the global redshifted 21~cm measurement.   The blade antenna, on the other hand, yields FoM values less than 1~mK over most of the sky, easily facilitating measurement of the global 21~cm signal.   The analytic \nicefrac[]{1}{2}-$\lambda$ wire antenna model produces the best results of all three antennas, with sub-mK residuals.  In all three cases, the FoM tends to reach its largest (worst) values when the Galactic plane and/or \textcolor{black}{Galactic Centre} is at or above the horizon. 
\subsection{Antenna Orientation}
For each antenna model and pointing we investigated two antenna orientations:  north-south (NS) aligned and east-west (EW) aligned excitation axis.  Orientation of the excitation axis is not a major factor if averaged over the entire latitude-LST range, but for a given latitude, one orientation may prove to have lower FoM values over a particular LST range. For the EDGES deployment, we find that a NS orientation is marginally superior at the deployment latitude of  $-26^\circ$ (see Fig. \ref{fig:pcolor_poly6_M}).
\subsection{Deployment Latitude}
Several sites have been considered and used for global redshifted 21~cm experiments.  Latitude $26^\circ$S is approximately the latitude of the two SKA sites (in South Africa and Australia).  EDGES and BIGHORNS are both currently deployed at the MRO, which is the SKA site in Western Australia.   DARE has tested prototype equipment at the MRO, as well as at the Greenbank observatory at latitude $38^\circ$N. SCI-HI is deployed at Guadalupe Island off the western coast of Mexico at 29$^{\circ}$2'N and is considering deployment at either Isla Socorro (18$^{\circ}$48'N) or Isla Clari\'{o}n (18$^{\circ}$22'N). Two other remote islands that are of interest to global 21 cm projects are Kiritimati (Christmas Island)  located 2000 km due south of Hawaii in the middle of the Pacific Ocean near the equator at 1$^{\circ}$52′N and Tristan da Cunha in the southern Atlantic Ocean between the southern tip of Africa and South America at 37$^{\circ}$4′S.

\begin{figure*}
 \begin{minipage}[b]{0.99\linewidth}
\centering
 \includegraphics{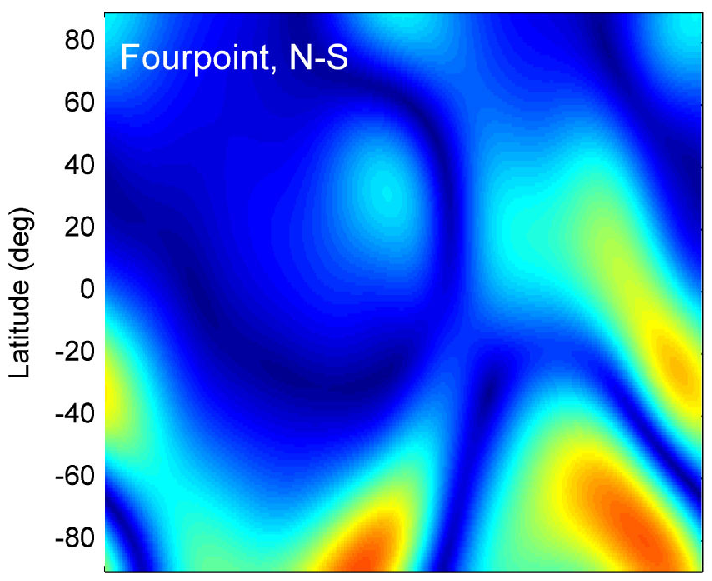}
 \includegraphics{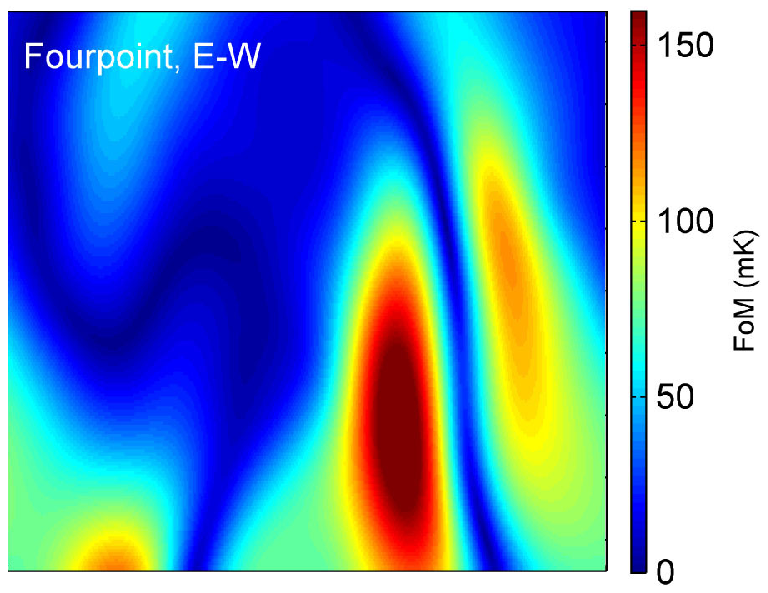}
  \end{minipage}
\begin{minipage}[b]{0.99\linewidth}
\centering
 \includegraphics{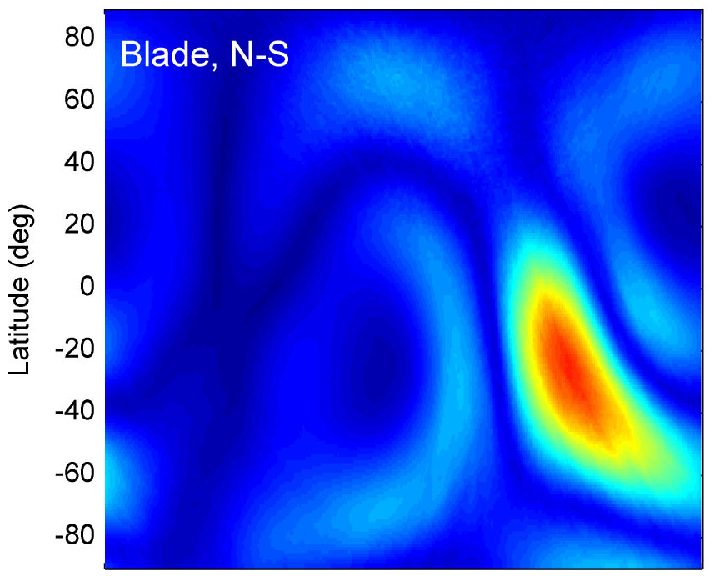}
 \includegraphics{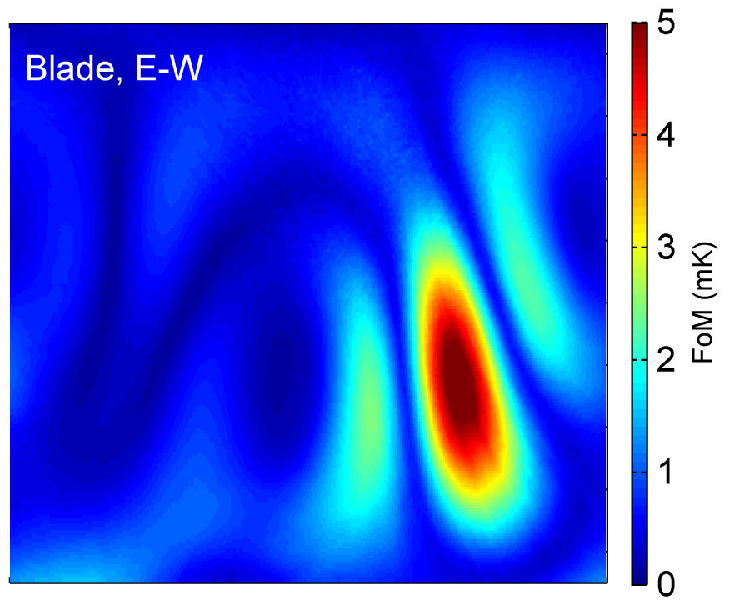}
 \end{minipage}
\begin{minipage}[b]{0.99\linewidth}
\centering
 \includegraphics{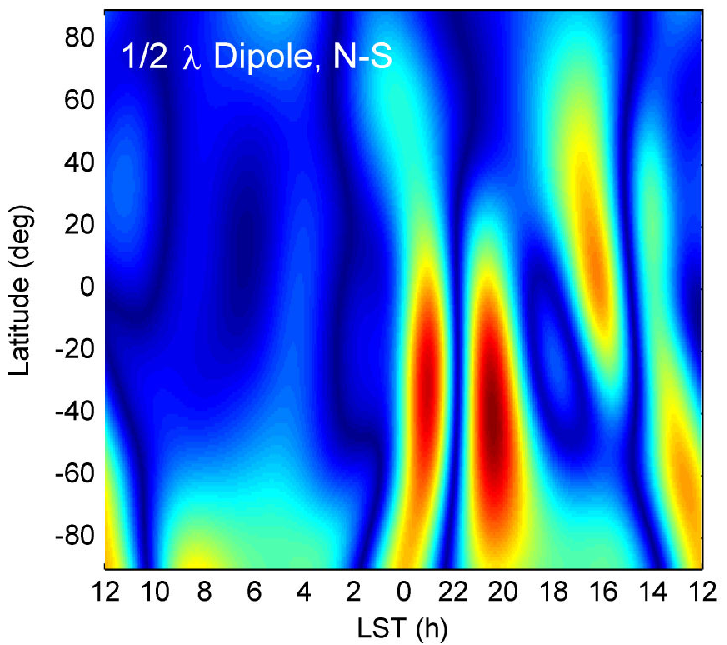}
 \includegraphics{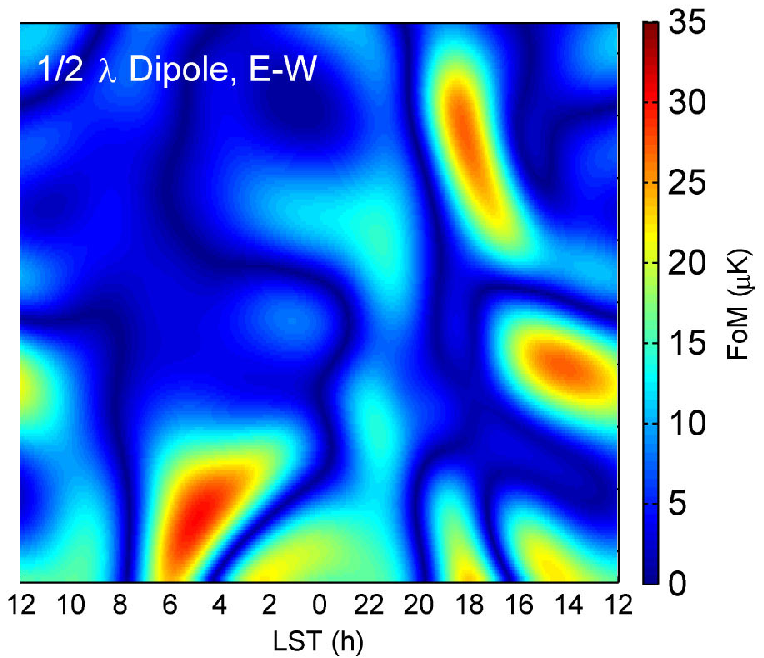}
 \caption{FoM as a function of latitude (declination) and LST (right ascension) for the three antennas.  From top to bottom, the panels show the fourpoint, blade, and analytic dipole models.  The excitation axis orientation is NS for the left column and EW for the right column. A six term polynomial was used for the fit.  The colour scale indicates the FoM magnitude and spans a different range for each antenna but is kept constant across the rows.  The fourpoint performs the worst and yields acceptably low FoM values (below $\sim$10~mK) only in a few areas of the sky (dark-blue regions in the top panel).  The blade antenna performs well with the FoM below 10~mK across the entire sky.  The \nicefrac[]{1}{2}-$\lambda$ analytic wire dipole model performs the best with sub-mK residuals.  The alignment does not greatly affect the distribution of FoM values, but for specific latitudes, one orientation may be better than another, suggesting that orientation choice must be evaluated for both the type of antenna used and the target deployment latitude. In all cases, the performance is best when the Galactic plane is below the horizon and generally worst when the Galactic plane and/or \textcolor{black}{Centre} (17~h 45~m LST, $-29^{\circ}$  dec) are visible near the horizon or at moderate zenith angles.}
\label{fig:pcolor_poly6_M}
 \end{minipage}
\end{figure*}

\begin{figure}
  \centering
  \includegraphics{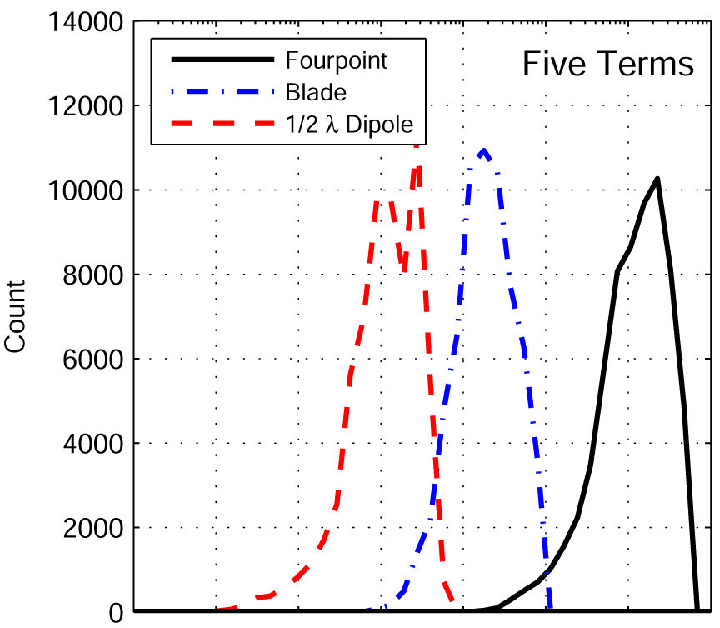}
  \includegraphics{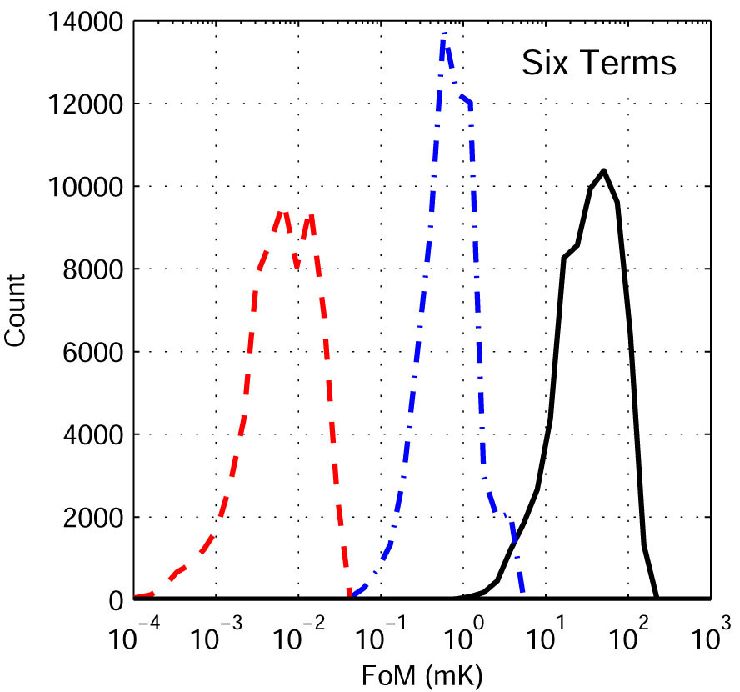}
  \caption{Plots of the FoM distribution for the fourpoint, blade, and analytic \nicefrac[]{1}{2}-$\lambda$ wire dipole antennas with a NS orientation for fits to (top) a five term polynomial and (bottom) a six term polynomial. The counts are the number of latitude and LST grid points that fall into FoM bins.  The trend towards lower FoM values with increasing polynomial terms is evident as well as the relative performance of the three antennas. \textcolor{black}{The fourpoint FoM is below 10~mK for 3\% of the data points using a 5 term fit and 12\% for a 6 term fit, while the blade is below 10~mK for 99.7 \% of the data points. The blade FoM is an order of magnitude better as the FoM is below 1~mK for 25\% of the data points using a 5 term fit and 73\% for a 6 term fit.}}
  \label{fig:FoM_histogram}
\end{figure}

The number of terms needed in the $T_\text{model}$ power-law polynomial fit to remove the chromatic beam and foreground to a certain RMS error level at a specific latitude and LST is dependent upon the type of antenna chosen. We have calculated the FoM for polynomials of length N=3 to 7 and have examined the results. The FoM values decrease as the number of terms in the polynomial increases, resulting in an increase in the number of  points with sub-mK level FoM values.  The histograms in  Fig. \ref{fig:FoM_histogram} quantify this trend for N=5 and N=6 for the three antennas. The counts are the number of latitude-LST grid points of the NS orientation of Fig.~\ref{fig:pcolor_poly6_M} that fall into FoM bins. The histogram plots indicate that the foreground can be removed in more Lat/LST locations with a given polynomial length using the blade antenna than using the fourpoint antenna.

\begin{figure}
  \centering
 \includegraphics{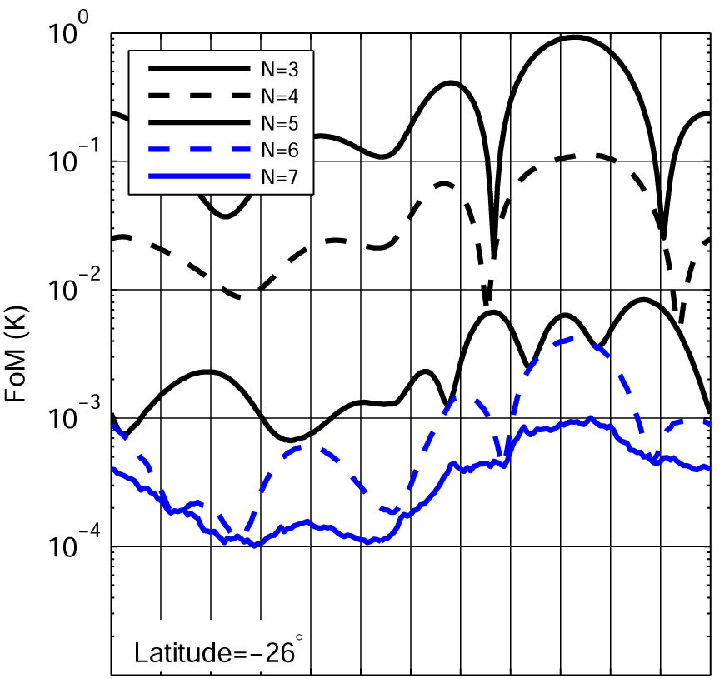}
  \includegraphics{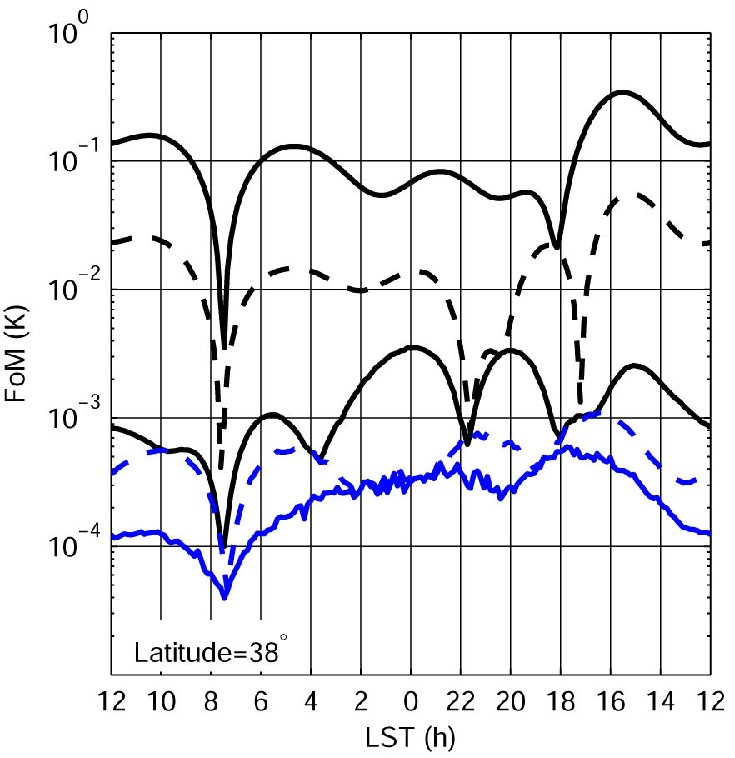}
  \caption{Blade FoM vs LST at latitude $-26^{\circ}$ (top) and latitude 38$^{\circ}$ (bottom), which correspond approximately to the latitude of the SKA sites in South Africa and Australia, as well as the EDGES fourpoint location, and to the latitude of the Greenbank Observatory respectively.  The antenna excitation axis is aligned NS.  The curves illustrate the effects of varying the number of polynomial terms in the $T_\text{model}$ fit, from N=3 (top) to N=7 (bottom).  For this antenna, acceptable FoMs were achieved with as few as 5 polynomial terms. The fourpoint and analytic \nicefrac[]{1}{2}-$\lambda$ wire antennas show similar progressions, but with different relative amplitudes.}
  \label{fig:blade_cut}
\end{figure}

For a given latitude of deployment, we take a cut through the plots of Fig. \ref{fig:pcolor_poly6_M} for the blade antenna at latitudes $26^\circ$S and $38^\circ$N with a NS aligned excitation axis and display the results in Fig.~\ref{fig:blade_cut}, which shows the FoM for the blade as a function of LST for polynomial lengths between N=3 and 7.  As expected, the FoM improves significantly as the number of polynomial terms is increased. At N=5, the FoM falls below the expected 21-cm signal strength across all LSTs.  The fourpoint and analytic \nicefrac[]{1}{2}-$\lambda$ wire antennas yield similar progressions, but with different overall amplitudes (not shown). Table \ref{tab:FoM} lists the FoM values for polynomials of length 3 through 7 terms for the antennas discussed at latitude  $-26^\circ$ at a relatively low FoM region near 4 h and at a relatively high FoM region near 17 h. For this latitude, Table~\ref{tab:FoM} indicates that the blade and fourpoint antenna are comparable for low values of N near regions of low FoM, but the blade performs better for higher values of N and especially for regions of high FoM. \textcolor{black}{All three antenna types listed can remove the sky foreground if six polynomial terms are used, and all but the fourpoint antenna are still acceptable with a 5 term polynomial. For 5-term fits and higher, the FoM performance between each of the three antennas improves by approximately an order of magnitude from the best-case \nicefrac[]{1}{2}-$\lambda$ dipole, to the blade, to the fourpoint design.}

\subsection{Global 21 cm Signal Detectability}
\textcolor{black}{While the above FoM anlysis establishes the relative performance between the antennas and for different pointings, we turn now to the SNR as defined in Eq. 7 to characterize the absolute performance of the antennas. Referring to Table \ref{tab:SNR}, a global 21~cm signal with a 20~MHz FWHM parameter is detectable with all 3 antennas using either 5 or 6 polynomial terms in the foreground fitting equation. However, a global 21~cm signal with a 40 MHz FWHM parameter is not detectable (SNR $<$ 2) for any of the antennas when a 6 term polynomial is used, given the assumed noise levels and the given latitude. The low FoM values of the blade antenna for a 5 term fit raises the SNR to 4.6 and 6.7, when the thermal noise is 3~mK and 2~mK respectively, and enables a global 21~cm signal detection even when it has a 40 MHz FWHM.}

\begin{table}
   \caption{FoM for polynomial lengths between 3 and 7 terms at latitude $-26^\circ$ and two LST values}
     \begin{tabular} {l   c c c c c c}
\hline
Antenna & \multicolumn{5}{c}{FoM (mK)  (LST = 4 h, $-26^\circ$)}  \\
 \multicolumn{1}{r}{Terms} &3&4&5&6&7\\
 \hline 
   Fourpoint                                              &   143 &19.5&8.20&3.93&3.86 \\
   Blade                                                   &   162 & 22.2&0.67&0.53&0.07\\
   \nicefrac[]{1}{2}-$\lambda$  wire dipole  &   3.45 & 0.16&0.06&0.01&$<$0.01\\
\hline
 & \multicolumn{5}{c}{FoM (mK)  (LST = 17 h, $-26^\circ$)}\\
 \hline
   Fourpoint                                             &   2200 &219 & 96.6 & 33.8 & 7.44 \\
   Blade                                                   &   999 & 121 & 6.60 & 3.88 & 0.53\\
   \nicefrac[]{1}{2}-$\lambda$  wire dipole &   118 & 7.60  & 0.24 & $<$0.01 & $<$0.01\\
   \hline
\end{tabular}
\label{tab:FoM}
\end{table}
\begin{table}
   \caption{\textcolor{black}{SNR for polynomial lengths between 3 and 7 terms at latitude $-26^\circ$ \textcolor{black}{with the Galatic Centre below the horizon}. Assumed spectral noise of 2-3~mK achieved by several nights of observational data averaging. One additional Gaussian 21~cm term added to the polynomial terms of the fit equation with FWHM of 20~MHz and 40~MHz.}}
\centering
     \begin{tabular}{lc@{\hspace{2.7em}}c@{\hspace{2.7em}}c@{\hspace{2.5em}}c@{\hspace{0.5em}}c@{}}
\hline
Antenna & \multicolumn{5}{c}{SNR  (FWHM = 20~MHz, Noise = 3~mK)}  \\
 \multicolumn{1}{r}{Poly Terms} &3&4&5&6&7\\
 \hline
   Fourpoint                                            &   0.5 & 2.4 & 7.8 & 5.3 & 5.4 \\
   Blade                                                  &   0.5 & 1.8 & 12 &7.0&7.0\\
   \nicefrac[]{1}{2}-$\lambda$  wire dipole &   14 & 13  & 12 &7.0 &7.0\\
   \hline
& \multicolumn{5}{c}{SNR  (FWHM = 40~MHz, Noise = 3~mK)}\\
 \hline
   Fourpoint                                             &   0.6 & 1.1 & 3.4 & 0.9 & 0.9 \\
   Blade                                                  &   0.6 & 0.8 & 4.6 & 1.1 & 1.1\\
   \nicefrac[]{1}{2}-$\lambda$  wire dipole &   13 & 5.1  & 4.7 &1.1 & 1.1\\
   \hline
& \multicolumn{5}{c}{SNR  (FWHM = 20~MHz, Noise = 2~mK)}\\
 \hline
   Fourpoint                                              &   0.5 &2.4&8.8&6.4&6.6 \\
   Blade                                                   &   0.5 & 1.8&17&11&11\\
   \nicefrac[]{1}{2}-$\lambda$  wire dipole  &   17& 19&18&11&11\\
\hline
 & \multicolumn{5}{c}{SNR  (FWHM = 40~MHz, Noise = 2~mK)}\\
 \hline
   Fourpoint                                             &   0.6 & 1.1 & 4.0 & 1.0 & 1.1 \\
   Blade                                                  &   0.6 & 0.8 & 6.7 & 1.6 & 1.7\\
   \nicefrac[]{1}{2}-$\lambda$  wire dipole &   18 & 7.7  & 7.0 &1.7 & 1.7\\
   \hline
\end{tabular}
\label{tab:SNR}
\end{table}
\subsection{\textcolor{black}{Spectral Derivative of Antenna Directivity}}
\label{subsec:origin}
\begin{figure*}
 \begin{minipage}{\textwidth}
\centering
 \includegraphics{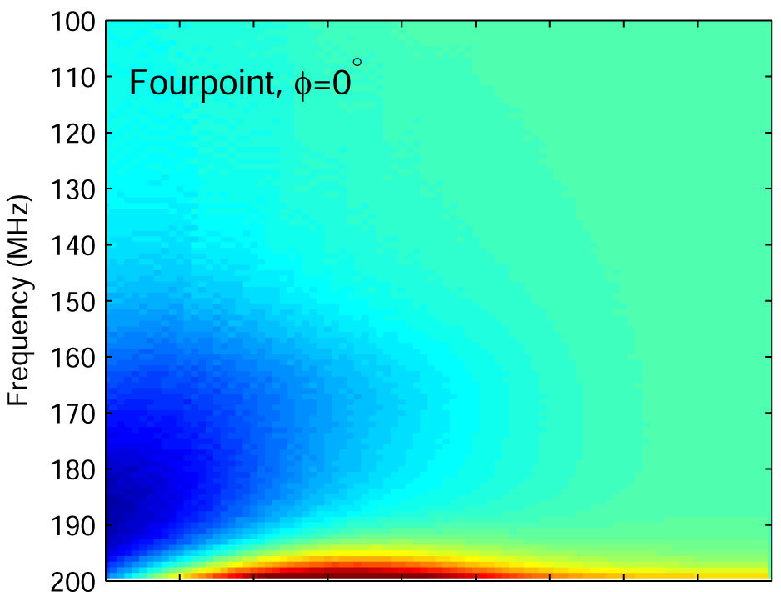}
 \includegraphics{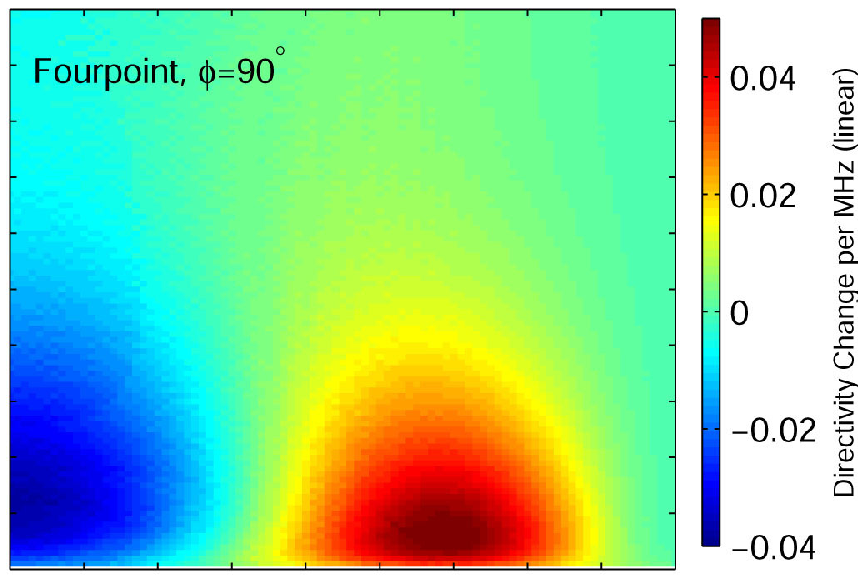}
  \end{minipage}
\begin{minipage}{\textwidth}
\centering
 \includegraphics{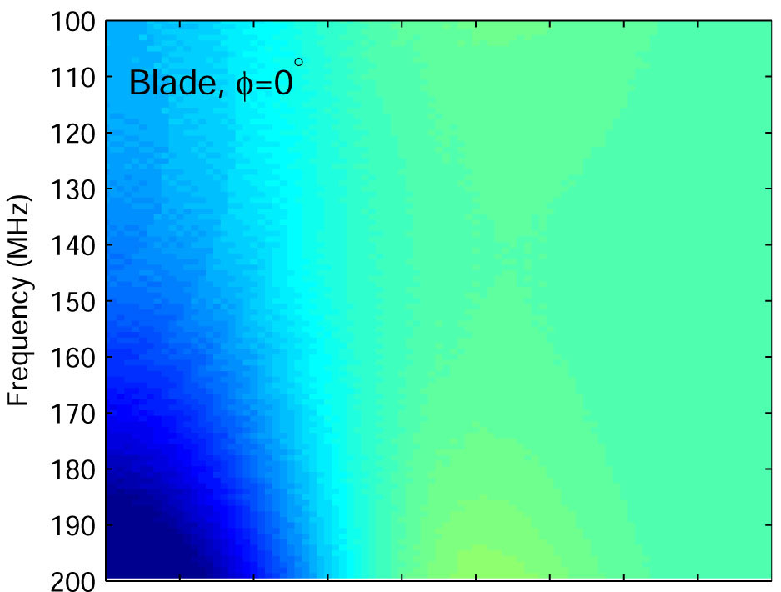}
 \includegraphics{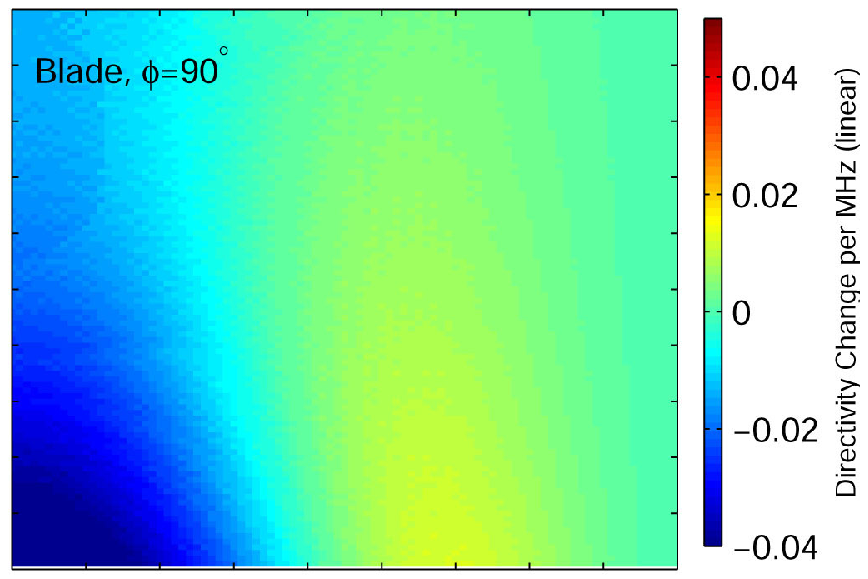}
 \end{minipage}
\begin{minipage}{\textwidth}
\centering
 \includegraphics{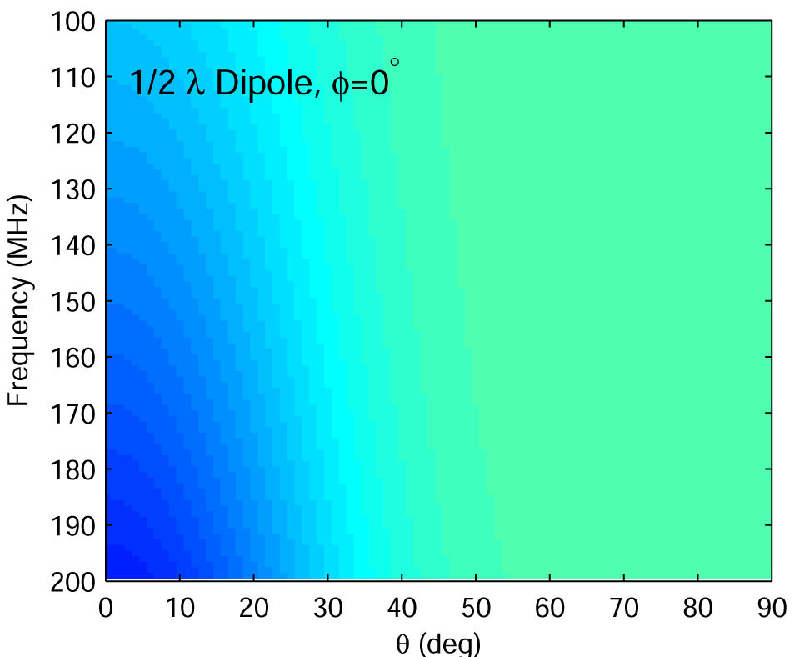}
 \includegraphics{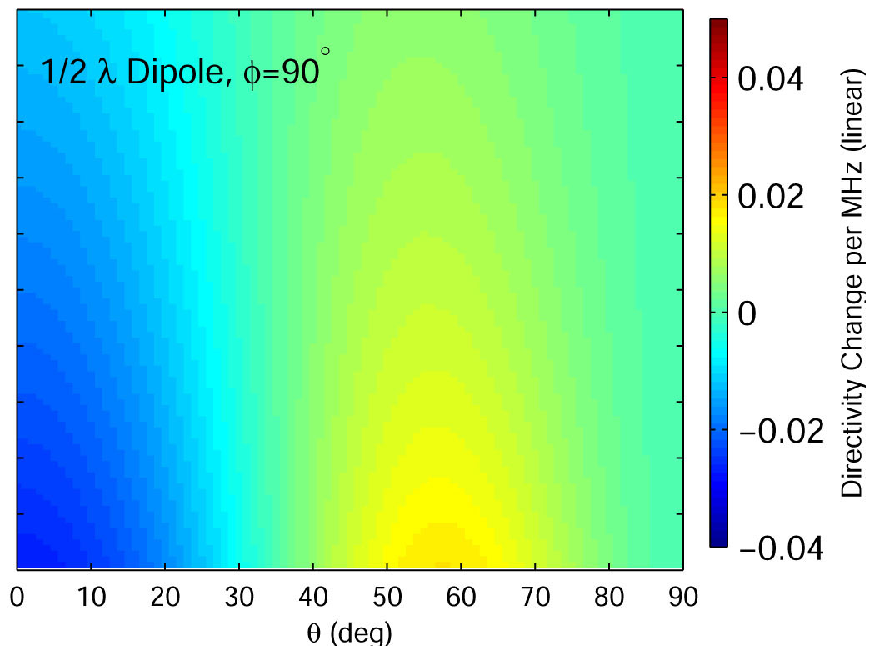}
 \caption{Derivatives of beam linear directivity values vs frequency at  (left) $\phi$=0$^\circ$ and (right) $\phi$=90$^\circ$ for the (top) fourpoint, (middle) blade, (bottom) analytic \nicefrac[]{1}{2}-$\lambda$ wire dipole antennas. Viewing these plots can quickly identify problems with the antenna beam that typical beam plots (see Fig. \ref{fig:cross-sections}) do not show.  All antennas show a decrease near the zenith with increasing frequency consistent with the beginning of structure due to the ground plane. The fourpoint antenna shows greater magnitude changes and additional structure both at the zenith and in other locations compared to the other two antennas. The blade antenna is more similar in amplitude and features to the analytic reference than the fourpoint antenna.}
\label{fig:Derivatives}
 \end{minipage}
 
\end{figure*}

During antenna design and simulation, visually examining beam plots of directivity vs elevation angle, $\theta$, \textcolor{black}{or even 3D plots at various frequencies} will not reveal chromatic issues with the beam, because the magnitude of the relevant beam features are on the order of 0.1 to 1.0 percent. \textcolor{black}{To examine the frequency structure at these small levels of change}, we take the derivative of the beam directivity with respect to frequency.  Figure \ref{fig:Derivatives} shows the beam derivative with respect to frequency vs. zenith angle for values of   $\phi$ at $0^\circ$ and $90^\circ$ for the three antennas. Excessive variation (rapid variation or multiple inflection points) in the beam derivative plot will indicate that the antenna will not perform well for 21 cm observations (high FoM values).

Referring to the plots in Fig. \ref{fig:Derivatives}, all antennas show a decrease near the zenith with increasing frequency, consistent with the beginning of structure due to  \textcolor{black}{ the height above the ground plane becoming a higher fraction of wavelengths} as discussed in section \ref{subsec:half_wavelength}. The fourpoint antenna shows greater magnitude changes and additional structure both at the zenith and in other locations compared to the other two antennas.  The blade antenna is more similar in amplitude and features to the analytic reference antenna than the fourpoint antenna, and correspondingly, the FoM values of the blade antenna are superior to those of the fourpoint antenna, i.e., the amount of beam structure correlates with FoM values. \textcolor{black}{The more complex shape of the fourpoint antenna, as compared to the blade antenna, may lead to more significant changes in the current flow pattern with frequency and consequently a larger change in beam shape with frequency.}

\section{CONCLUSION}
\label{sec:conclusion}

We \textcolor{black}{evaluated two dipole-based antennas used by EDGES and one idealized reference antenna} to assess the effects of frequency-dependent beam shapes on the ability to remove the foreground from global redshifted 21~cm measurements \textcolor{black}{and detect the redshifted global 21~cm signal}.  \textcolor{black}{Across the full latitude-LST space we found that the fourpoint antenna produced sub 10~mK FoM values in 3\% and 12\% of the locations for \textcolor{black}{foreground fits using polynomials with 5 and 6 terms respectively}, while the \textcolor{black}{FoM values of the blade antenna were} below 10~mK in over 99\% of the locations for both fits. Furthermore, \textcolor{black}{FoM values of foreground fitting for the blade antenna were} below 1~mK in 25 \% and 72 \% of the locations for 5 and 6 term fits respectively. We note that the optimum choice of E-W or N-S excitation axis orientation depends upon specific deployment location as one orientation was not always better than the other.} The fourpoint antenna is only suitable at a few restricted locations on the sky using a 5 or 6 term fit, while the blade antenna provides adequate FoM performance across the entire sky when using a \textcolor{black}{5 or} 6 term polynomial to remove the foreground.

\textcolor{black}{In our simulations, a narrow 21 cm signal corresponding to a rapid reionization over 20 MHz was detectable for all antennas assuming 3~mK of thermal noise. The SNR values indicate detection is favorable for either a 5 or 6 term foreground removal fit. For a 5 term fit, the \textcolor{black}{SNR values of the} blade and analytical dipole are nearly the same ranging from 12 to 18, and the fourpoint \textcolor{black}{values range} between 7.8 and 8.8. For a 6 term fit, the \textcolor{black}{SNR values of the} blade and analytical dipole are the same ranging from 7 to 11, and the fourpoint \textcolor{black}{values range} between 5.3 and 6.4 depending upon the thermal noise.}

\textcolor{black}{A slower reionization over 40 MHz is not detectable with any of our antennas when the foreground is fitted with a 6 term polynomial as the SNR is no greater than 1.7.  When a five term polynomial is used the SNR increases and the detection is again favorable. The SNR for the blade antenna is between 4.6 and 6.7, between 3.4 and 4.0 for the fourpoint, and between 4.7 and 7.0 for the analytical dipole, again depending upon thermal noise.  Based upon this ananlysis we conclude that the blade antenna using a five term polynomial with thermal noise averaged down to $<$ 3~mK is capable of detecting or placing meaningful limits on the global 21 cm signal during reionization.}

During antenna design the beam derivative with respect to frequency plot is a convenient tool to quickly assess the frequency structure in the beam and thus the ensuing effectiveness of foreground removal. This method can reveal problems quickly and requires little computing power.

Although we studied the frequency range 100 - 190~MHz, the results we have reported can be applied to other frequency ranges since the properties of an antenna scale linearly in wavelength with the physical size of the antenna.  For example, global 21~cm experiments targeting the First Light signal between 50 and 100~MHz can also use these results by scaling the antenna design by a factor of two and halving the frequency. The FoM scales as $\nu^{-2.5}$. 

A variety of non-dipole antenna designs have been considered for global 21~cm experiments, such as log-dipole and horn antennas.   Most of these antennas have considerably more structure than simple dipole antennas and can be expected to exhibit even larger chromatic effects.  Detailed investigation of other antennas is left for future work.
\section*{Acknowledgments}

This work was supported by the NSF through research awards for the Experiment to Detect the Global EoR Signature (AST-0905990 and AST-1207761) and by NASA through Cooperative Agreements for the Lunar University Network for Astrophysics (NNA09DB30A) and the Nancy Grace Roman Technology Fellowship (NNX12AI17G). EDGES is located at the Murchison Radio-astronomy Observatory. We acknowledge the Wajarri Yamatji people as the traditional owners of the Observatory site.

\label{lastpage}

\bsp

\end{document}